\documentclass{aa}
\usepackage{psfig}

\def\bron{SAX J1819.3--2525}
\def\optic{V4641~Sgr}
\def\ecs{erg~cm$^{-2}$s$^{-1}$}
\def\pcs{ct~cm$^{-2}$s$^{-1}$}
\def\lum{erg~s$^{-1}$}

\begin{document}
\thesaurus{06(02.01.2; 08.02.1; 08.09.2 \bron, XTE~J1819--254, \optic; 13.25.5)}

\title{BeppoSAX observations of the nearby low-mass X-ray 
binary and fast transient \bron}
\author{J.J.M.~in~'t~Zand\inst{1}
 \and E.~Kuulkers\inst{1,2}
 \and A.~Bazzano\inst{3}
 \and R.~Cornelisse\inst{1,2}
 \and M.~Cocchi\inst{3}
 \and J.~Heise\inst{1}
 \and J.M.~Muller\inst{1,4}
 \and L.~Natalucci\inst{3}
 \and M.J.S.~Smith\inst{1,5}
 \and P.~Ubertini\inst{3}
}
\offprints{J.J.M.~in~'t Zand (at e-mail {\tt jeanz@sron.nl})}

\institute{     Space Research Organization Netherlands, Sorbonnelaan 2,
                NL - 3584 CA Utrecht, the Netherlands
	 \and
		Astronomical Institute, Utrecht University, P.O. Box 80000,
		NL - 3508 TA Utrecht, the Netherlands
         \and
                Istituto di Astrofisica Spaziale (CNR), Area Ricerca Roma Tor
                Vergata, Via del Fosso del Cavaliere, I - 00133 Roma, Italy
         \and
                BeppoSAX Science Data Center, Nuova Telespazio,
                Via Corcolle 19, I - 00131 Roma, Italy
         \and
                BeppoSAX Science Operation Center, Nuova Telespazio,
                Via Corcolle 19, I - 00131 Roma, Italy
                        }
\date{Received, accepted }

\authorrunning{In 't Zand et al.}
\titlerunning{BeppoSAX observations of \bron}

\maketitle

\begin{abstract}
\bron\ is a nearby X-ray transient which exhibited a fast and large
X-ray outburst
on Sep. 15, 1999 (Smith et al. 1999). The Wide Field Cameras and the 
Narrow Field Instruments (NFI) on board {\em BeppoSAX\/} observed \bron\ 
at various stages of its activity before that, in the spring and fall of 
1999. The fluxes 
range between 0.012 and 0.3 Crab units (2-10 keV). The NFI observation is 
unique because 
it is the longest semi-continuous observation of \bron\ so far, and it offers 
a study of the spectrum at
a relatively high resolution of 8\% full width at half maximum at 6~keV.
We discuss the observations with emphasis on the X-ray spectrum.
A strong Fe-K emission line was detected in \bron\ with an equivalent 
width between 0.3 and 1~keV. The line energy is up to 6.85$\pm$0.02~keV
and suggests the presence of
highly ionized iron.
We identify this as fluorescent emission from a photo-ionized plasma.
The continuum spectrum is typical for a low-mass X-ray binary in which
emission from an accretion disk corona plays an important role.
There is no sign of an eclipse or periodic signal due to the binary orbit 
in this exposure, despite the fact that the twin jets seen at radio 
wavelengths suggest a high inclination angle.
\end{abstract}

\section{Introduction}
\label{intro}
The X-ray source \bron\ was discovered in February,
1999, independently with the Wide Field Cameras (WFCs) on {\em BeppoSAX\/}
(In 't Zand et al. 1999a) and with the Proportional Counter Array (PCA) on
{\em RossiXTE\/} (ergo, its alternative designation XTE~J1819--254,
Markwardt et al. 1999a). The WFC detection involved an hour-long flare 
with a peak of 80~mCrab on Feb. 20. The first PCA detection 
occurred during regular scans of the Galactic Center field, on Feb. 18.

For some time there was confusion over the identification of the optical
counterpart. The variable star GM~Sgr is within the 2\arcmin-radius WFC error 
box of the X-ray
source as noted by In~'t~Zand et al. (1999a). However, it was later 
discovered that the actual counterpart is another, previously unknown,
variable star only $\sim$1\arcmin\ north of GM~Sgr: \optic\
(Williams 1999, Samus et al. 1999).

\optic\ was found to show a large outburst in visual magnitude $m_{\rm V}$ 
from about 11 on Sep. 14.8 up to 8.8 on Sep. 15.4 (Stubbings 1999).
X-ray observations with the All-Sky Monitor on {\rm RossiXTE\/} also
showed a bright outburst of \bron\ up to a peak intensity of 
12.2~Crab on Sep. 15.7 (2-12 keV, Smith et al. 1999). The giant X-ray
outburst was preceded by a 4.5~Crab precursor 0.8~d earlier. Between
both peaks the flux decreased to a quiescent level. Wijnands \& Van der
Klis (2000) report that during the tail of the giant X-ray outburst, 
strong variability was observed of factor-of-four on seconds time scales
to factor-of-500 on minutes time scales. The giant outburst
was also seen at radio wavelengths (Hjellming et al. 1999a, 1999b).
The radio source was resolved and had the appearance of a twin jets 
structure extending 0.25\arcsec\ each jet. The giant and precursor 
outburst were, furthermore, seen in 20-100 keV (McCollough et al. 1999). 

Within 2~d after the giant outburst, \optic\ settled down to a quiescent
brightness of $m_{\rm V}=13.5$. Combined with the failure 
of any X-ray detection
since the giant outburst (C.B. Markwardt, priv. comm.), this shows that
the flaring activity faded considerably, although the Balmer lines 
continued to show additional activity up to at least Sep. 30 (Wagner 1999). If 
one assumes that the emission is due to a main sequence A star (Garcia et al.
1999), one obtains
with B--V=--0.2 and E(B--V)=0.24 (Wagner 1999) a distance of 0.4~kpc. 
We will adhere to this distance in the present paper but acknowledge its
uncertainty. From the magnitudes alone, we estimate the uncertainty to be 
about 50\%. With a 0.4~kpc distance, the giant outburst would have a peak 
X-ray luminosity of 
$5\times10^{36}$~\lum\ in 2-10 keV. This luminosity, and the spectrum
discussed in the present paper, clearly suggest 
that there is a compact object at play that is either a neutron star 
or black hole. It also identifies the object as a low-mass X-ray binary 
(LMXB), and the short distance implies it is possibly the {\em nearest} 
LMXB known
so far (c.f., Van Paradijs \& White 1995). This opens up interesting 
perspectives
for \bron\ as a case study of the quiescent emission from transient 
low-mass X-ray binaries. We note that recently Rutledge et al. (1999) 
introduced diagnostics for the nature of the compact object from the quiescent 
emission.

In this paper, we present the observations of \bron\ with the WFCs and
Narrow Field Instruments (NFI) on {\em BeppoSAX\/} in the spring and fall 
of 1999. This includes the longest semi-continuous observation of the system
so far. \bron\ was observed in two states: a flaring state with peak fluxes 
that are one to two orders of magnitude less than that of the giant outburst,
and a 'calm' state with an average flux that is about three orders of
magnitude smaller (though still considerably higher than that of the
quiescent emission).

\begin{figure}[t]
\psfig{figure=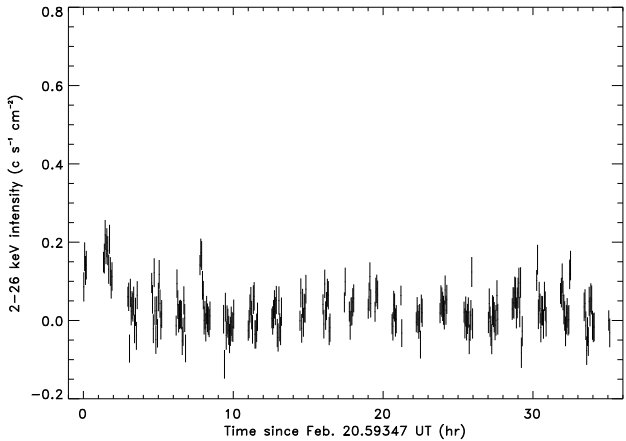,width=\columnwidth,clip=t}

\caption[]{Light curve of first flare on Feb. 20, as measured with the WFC 
unit 1 in the 2 to 26 keV band. The time resolution is 200~s.
\label{figwfclc1}}
\end{figure}

\begin{figure}[t]
\psfig{figure=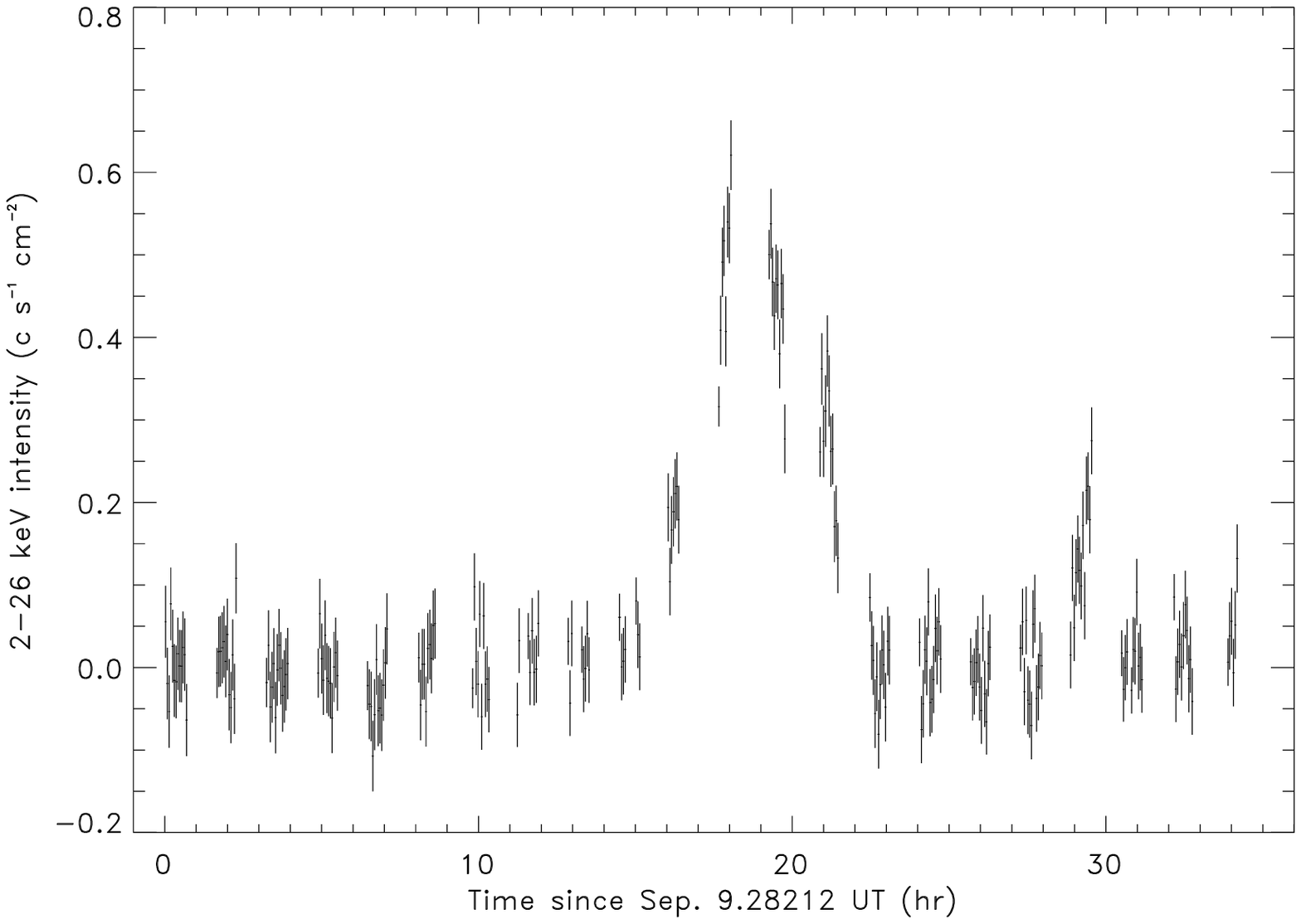,width=\columnwidth,clip=t}

\caption[]{Light curve of first flare on Sep. 10, as measured with the WFC 
unit 2 in the 2 to 26 keV band. The time resolution is 200~s.
\label{figwfclc2}}
\end{figure}

\begin{figure}[t]
\psfig{figure=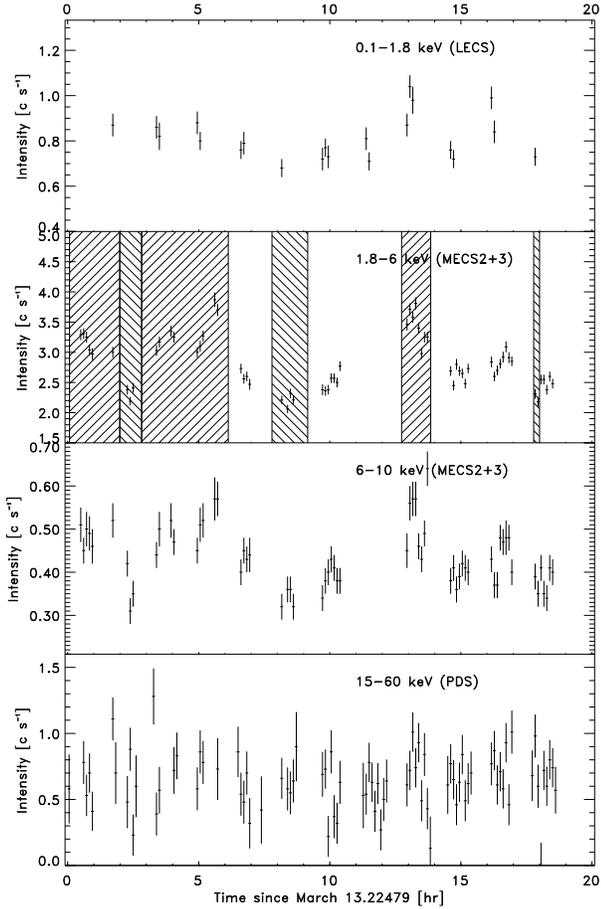,width=\columnwidth,clip=t}

\caption[]{Light curve as measured with the NFI in a number of bandpasses, 
corrected for background. The time resolution is 400~s. In the second panel
from above the time intervals have been hatched that refer to low-flux
intervals ($\backslash\backslash$ hatched) or high-flux intervals 
($//$ hatched).
\label{fignfilc}}
\end{figure}

\section{Observations}
\label{secobs}

The {\em BeppoSAX\/} platform carries two sets of astrophysical X-ray and 
$\gamma$-ray devices in space (Boella et al. 1997a). One pertains to two
identical Wide Field Cameras (WFCs, Jager et al. 1997) that view the sky with 
$40\times40$ square degrees field-of view in opposite directions with
5\arcmin\ spatial resolution in the 2 to 26 keV bandpass. The other set
includes the Narrow Field Instruments (NFI) that are co-aligned in a direction
that is perpendicular to that of both WFCs.
The NFI include 2 imaging instruments that are active below 10~keV,
the Low-Energy and the Medium-Energy Concentrator Spectrometer (LECS and
MECS respectively, see Parmar et al. 1997 and Boella et al. 1997b 
respectively), and two non-imaging instruments that have bandpasses of
$\sim12$ to 300~keV (the Phoswich Detector System or PDS,
Frontera et al. 1997) and 4 to 120~keV (the High-Pressure Gas 
Scintillation Proportional Counter or HP-GSPC, Manzo et al. 1997). 
The MECS has a photon energy resolution of 8\% (full width at half maximum)
at 6~keV.

Since mid 1996, the WFCs carry out a long-term program of monitoring 
observations in the field around the Galactic center.
The program consists of campaigns during the spring and fall of each year. 
Each campaign lasts about two months and typically comprises weekly
observations. \bron\ was detected twice during these campaigns so far, 
in hourly exposures above a $\sim$50~mCrab detection limit, on Feb. 20 
and Sep. 10, 1999. The first WFC-detection triggered a 
target-of-opportunity observation (TOO) with the NFI on March 13.22-14.02, 
1999 (this is 186~d before the giant outburst). \bron\ was strongly detected 
in three NFI (the HP-GSPC was not turned on), and the average intensity was 
found to be about 12~mCrab (2-10 keV). The NFI net exposure times on \bron\
are 10.2~ks for LECS, 27.7~ks for MECS and 14.7~ks for PDS. The LECS and MECS 
images show a single bright source, the position as determined from the MECS 
image is R.A.~18$^{\rm h}$19$^{\rm m}$22\farcs2, 
Decl.~=~--25\degr24\arcmin03\arcsec\ (Eq.~2000.0, error radius 0\farcm8) which
is 1\farcm0 from that determined with the WFC (In~'t~Zand et al. 1999a) and
0\farcm4 from the optical counterpart \optic.

\section{Light curves}
\label{seclc}

The background-subtracted light curves of the two flares as measured with 
the WFCs are presented 
in Figs.~\ref{figwfclc1} and \ref{figwfclc2}. Besides these two occasions,
the source was never detected in either short 1-hr or long 24-hr WFC 
exposures with typical upper limits of 50 and 12~mCrab respectively. The 
light curves of both detections are characterized by sporadic flaring 
activity up to 0.2~\pcs, which corresponds 
to 0.1~Crab units, with a larger and longer flare on Sep. 10 with a peak of 
0.3~Crab units. The large flare was above the detection limit for about 6~hr 
and occurred in the middle of a 34~hr long observation. It started on Sep. 
10.0, which is 5.4~d before the giant outburst. Its duration is comparable
to that of the giant flare.

The flaring activity on a time scale of hours is in line with the behavior 
as measured with the Proportional Counter Array (PCA) on {\em RossiXTE}
(Markwardt et al. 1999b). The 
PCA covers \bron\ bi-weekly since Feb. 5, 1999, for individual snapshot 
exposures of about 60~s with a sensitivity
of about 1~mCrab. \bron\ was first seen on Feb. 18 and subsequently 
showed an erratic variable behavior until immediately after the giant 
outburst, with fluxes ranging from 0.5 to 30~mCrab (Markwardt et al. 1999b).
Since the giant outburst, no emission was detected anymore (C.B.~Markwardt, 
priv. comm.).

The WFC data do not reveal any type I X-ray burst from \bron\ in about 0.8~Ms
of source coverage for the year 1999 when it was seen to be active with
the PCA (Markwardt et al. 1999b), or for 
$\sim$3~Ms over all WFC observations since 1996.

Fig.~\ref{fignfilc} shows the evolution of the background-corrected photon 
count rates in various bandpasses of the three NFI, in 400~s time resolution. 
For an explanation of the method of data extraction, we refer to In~'t~Zand 
et al. (1999b). The light curves show slow variability on a time scale of 
a few hours with an amplitude of about 50\%. The average flux level is
12~mCrab (2-10 keV) which, relative to the flares, can be regarded as
calm emission. However, we note that this cannot be regarded as
quiescent emission because the PCA has seen flux levels from this source
at least one order of magnitude smaller (Markwardt et al. 1999b).

A power density spectrum of the MECS photon count rate, generated with
a timing resolution of 0.02~s and averaged over 256~s time intervals, 
reveals no measurable narrow features nor broad-band noise. The upper 
limit on the variability, integrated between 0.01 and 10~Hz with
power laws with indices of --0.5, --1.0 and --2.0, are 18, 8 and 3.5\% 
fractional rms
respectively (90\% confidence). The upper limit on the pulsed amplitude
is 1.3\% fractional rms (95\% confidence) in 0 to 25 Hz.

\begin{table}[tb]
\caption[]{Best-fit parameter values of the Comptonized model to the NFI 
spectrum. The last line specifies the $\chi^2_{\rm r}$ value for the fit 
without a bb (black body) component. This value applies
after re-fitting the remaining parameters. EW means equivalent width.}
\begin{tabular}{ll}
\hline
Model              & Comptonized + 2 narrow lines +\\
                   & black body\\
$N_{\rm H}$        & $(0.05\pm0.02)\times10^{22}$~cm$^{-2}$ \\
bb $kT$            & $1.12\pm0.09$ keV\\
bb $R/d_{\rm 10~kpc}$ & $1.8\pm0.3$ km \\
Wien $kT_{\rm W}$  & $0.36\pm0.02$ keV \\
Plasma $kT_{\rm e}$& $17\pm6$ keV\\
Plasma optical     & $1.5\pm0.6$ for disk geometry \\
\hspace{2mm}depth $\tau$       & $3.6\pm0.9$ for spherical geometry \\
Comptonization     &  $0.3\pm0.1$ for disk geometry \\
\hspace{2mm}parameter $y$      &  $1.7\pm0.4$ for spherical geometry\\
Emission line energies & 6.68 and 6.96 ~keV (fixed)\\
Emission line widths & 0~keV (fixed)\\
Flux line 6.68~keV & $(2.8\pm0.9)\times10^{-4}$~phot~s$^{-1}$cm$^{-2}$ \\
Flux line 6.96~keV & $(3.8\pm0.9)\times10^{-4}$~phot~s$^{-1}$cm$^{-2}$ \\
EW line 6.68~keV   & 102 eV\\
EW line 6.96~keV   & 164 eV\\
Error in combined EW& 30 eV \\
$\chi^2_{\rm r}$   & 1.30 (96 dof) \\
Flux (2-10 keV)    & $2.7\times10^{-10}$~erg~s$^{-1}$cm$^{-2}$ \\
Flux (0.4-120 keV) & $5.4\times10^{-10}$~erg~s$^{-1}$cm$^{-2}$ \\
$\chi^2_{\rm r}$ without bb  & 2.06 (98 dof) \\
\hline
\end{tabular}
\label{tabnfifit}
\end{table}

\begin{figure}[t]
\psfig{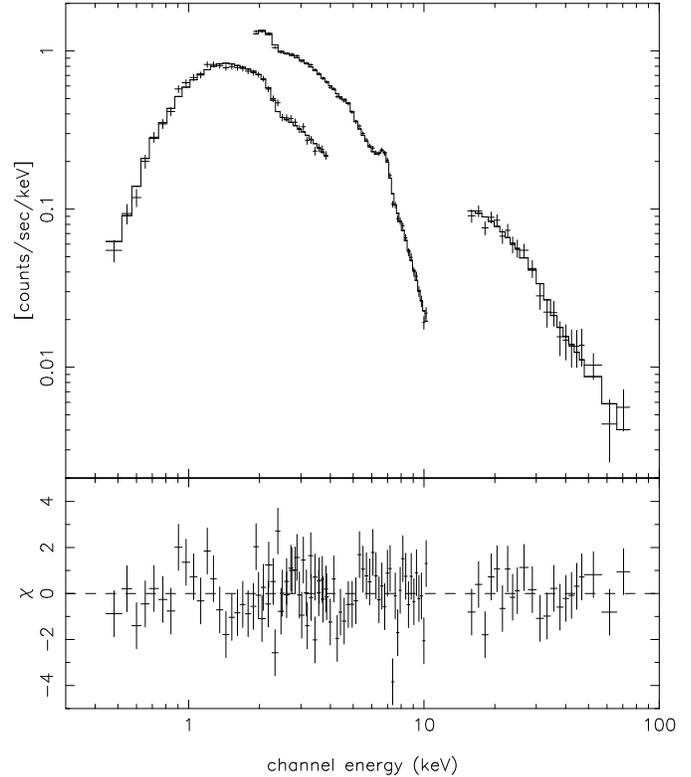}

\caption[]{Upper panel: count rate spectrum (crosses) and Comptonized spectrum
model (histogram) for the average emission. Lower panel: residual in units
of sigma per channel.
\label{fignfispectrum}}
\end{figure}

\section{Spectrum of calm emission}
\label{secnfispec}

\begin{figure}[t]
\psfig{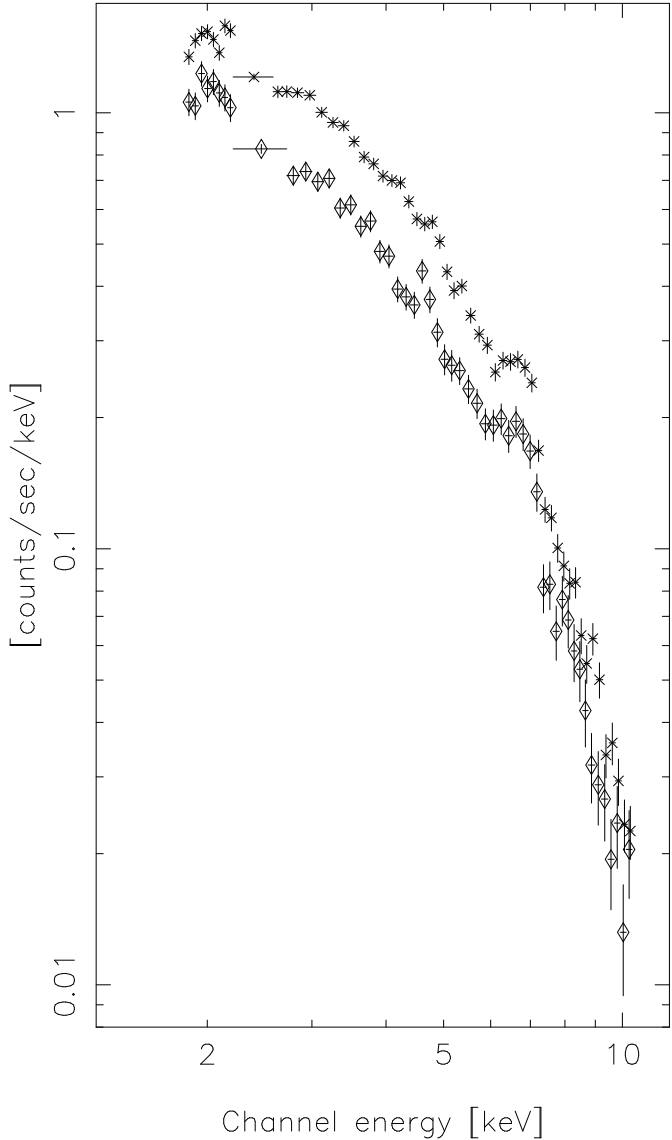}

\caption[]{Average MECS spectrum for the low-flux (lower) and high-flux
intervals (upper) of the data (see also Fig.~\ref{fignfilc}).
\label{fig2mecsspectra}}
\end{figure}

\begin{figure}[t]
\psfig{figure=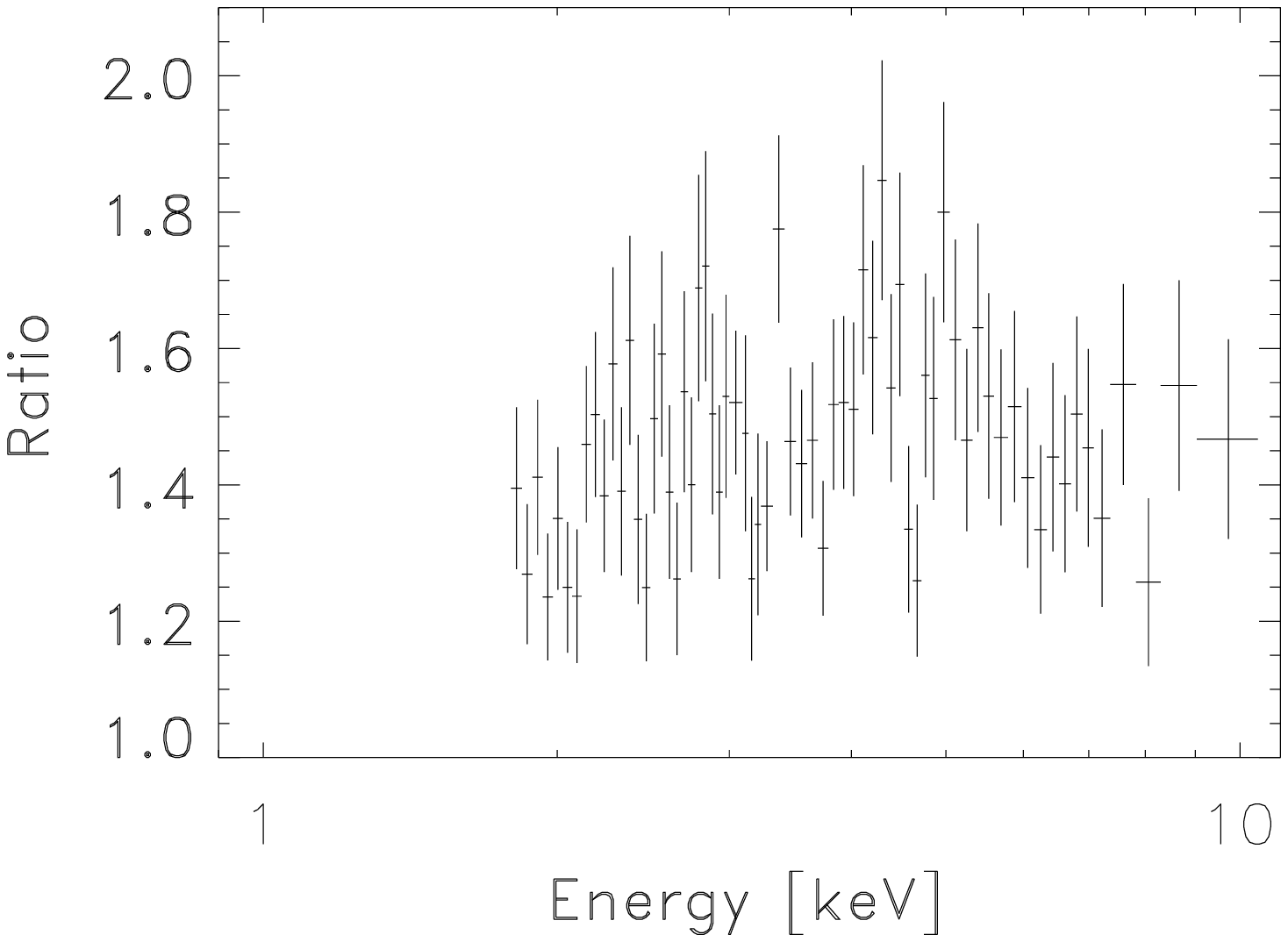,width=\columnwidth,clip=t}

\caption[]{Ratio of count rate spectrum between bright and
faint selections of NFI data. The binning was arranged such that
the error per data point is less than 10\%.
\label{fignfiratio}}
\end{figure}

We first extracted a spectrum for the average emission over the whole NFI
observation. The spectral channels were rebinned so as to
sample the spectral full-width at half-maximum resolution by three bins
and to accumulate at least 20 photons per bin. The bandpasses were limited 
to 0.4-4.0 keV (LECS), 1.8-10.5~keV (MECS) and 
15-120 keV (PDS) to avoid photon energies where either the spectral 
calibration of the instruments is not complete or no flux was measured 
above the statistical noise. We tried to model the spectrum with various 
descriptions. In all models, an allowance was made to leave free - within 
reasonable limits - the relative normalization of the spectra from LECS and
PDS
to that of the MECS spectrum, to accommodate cross-calibration uncertainties
in this respect. Publicly available instrument response functions and software
were used (version November 1998). 

The continuum could best be fitted ($\chi^2_{\rm r}=1.30$ for 96 dof)
with a Comptonization model
(Titarchuk 1994) plus black body radiation, see Table~\ref{tabnfifit}. 
Next to that there is a strong emission line at 7~keV. A fit with
a single narrow line results in a centroid energy of 6.85$\pm0.02$~keV. 
We identify this as K$\alpha$ fluorescence in strongly ionized iron.
The centroid energy is between the expected Fe-K
lines for helium-like (6.68 keV) and hydrogen-like Fe (6.96~keV).
We included in the model narrow lines at these fixed energies. 
The best-fit parameter values are
given in Table~\ref{tabnfifit} and a graph of the spectrum and
the model fit in Fig.~\ref{fignfispectrum}. The fits with other continuum
models, in combination with a black body and two narrow line components
results in fit qualities of $\chi^2_{\rm r}=3.19$ for 98 dof (thermal
bremsstrahlung), $\chi^2_{\rm r}=1.83$ for 96 dof (broken power law)
and $\chi^2_{\rm r}=1.59$ for 96 dof (power law with high-energy cut off).

Terada et al. (1999) reported about the 1996 transient AX~J1842.8--0423
which exhibited an Fe-K line at 6.80~keV with a large equivalent 
width of 4~keV. The 0.5-10~keV continuum plus line spectrum was
successfully fitted with a thin hot thermal plasma emission model of
temperature 5.1~keV. We fitted such a model according to the MEKAL code
implementation (Mewe et al. 1995).
The fit was reasonable, provided two additional components were
included. With a power law and black body as additional
components, $\chi^2_{\rm r}=1.73$ (98 dof), which is a worse fit 
than the Comptonization model in Table~\ref{tabnfifit}. The
resulting plasma temperature is $9.1\pm0.5$~keV. The fitted contribution 
of the thin plasma to the flux is of order 10\%. The emission measure
of the thermal plasma is 
$(1.33\pm0.08)\times10^{56}(d/0.4~{\rm kpc})^2$~cm$^{-3}$.

E(B--V)=+0.24 (Wagner 1999) implies $N_{\rm H}=0.13\times10^{22}$~cm$^{-2}$
according to the relationship defined by Predehl \& Schmitt (1995). If we 
assume that the uncertainty in E(B--V) is 0.10, where most of the uncertainty 
comes from the uncertainty in the calibration of the relationship used by
Wagner (1999) between the
equivalent width of the 578.0~nm interstellar absorption line to E(B--V) (see 
Herbig 1975), then the error in $N_{\rm H}$ is 0.05$\times10^{22}$~cm$^{-2}$.
If we fix $N_{\rm H}$ to $0.08\times10^{22}$~cm$^{-2}$ and leave free the 
remaining parameters of the Comptonized model in Table~\ref{tabnfifit}, 
$\chi^2_{\rm r}$ is 1.31 (97 dof). We conclude that $N_{\rm H}$ 
as determined from the X-ray spectrum is consistent with 
E(B--V)=+0.24$\pm0.10$.

To determine whether the variability as illustrated in Fig.~\ref{fignfilc}
is accompanied by strong spectral changes, we extracted a spectrum for
times when the source was relatively faint and one for times when the 
source was relatively bright. These times are indicated by hatched areas
in Fig.~\ref{fignfilc}. Subsequently, we employed the same Comptonized model
as for the whole observation, leaving free only the normalizations of the
different contributions. The resulting values for $\chi^2_{\rm r}$ are
1.36 for the bright data (104 dof) and 1.22 (103 dof) for the faint data.
The 44\% difference in the 0.4-10 keV flux between the faint and bright data is
due to approximately equal changes in blackbody and Comptonized components 
(i.e., 25 and 19\% respectively). The flux of the emission lines scales with
the integral flux: the combined equivalent widths of both lines is 
identical in both cases at 0.26 keV. This spectral behavior is illustrated in 
Fig.~\ref{fig2mecsspectra} which zooms in on the MECS
part of the spectrum  (including the emission lines) for the two extremes.
In Fig.~\ref{fignfiratio} the ratio between both spectra is presented.
This is consistent with a constant ratio of $1.44\pm0.02$ throughout the 
spectrum ($\chi^2_{\rm r}=1.18$ for 65 dof). The apparent bump between 4 
and 5~keV is statistically not significant.

\section{Spectrum of flare}
\label{secwfcspe}

\begin{figure}[t]
\psfig{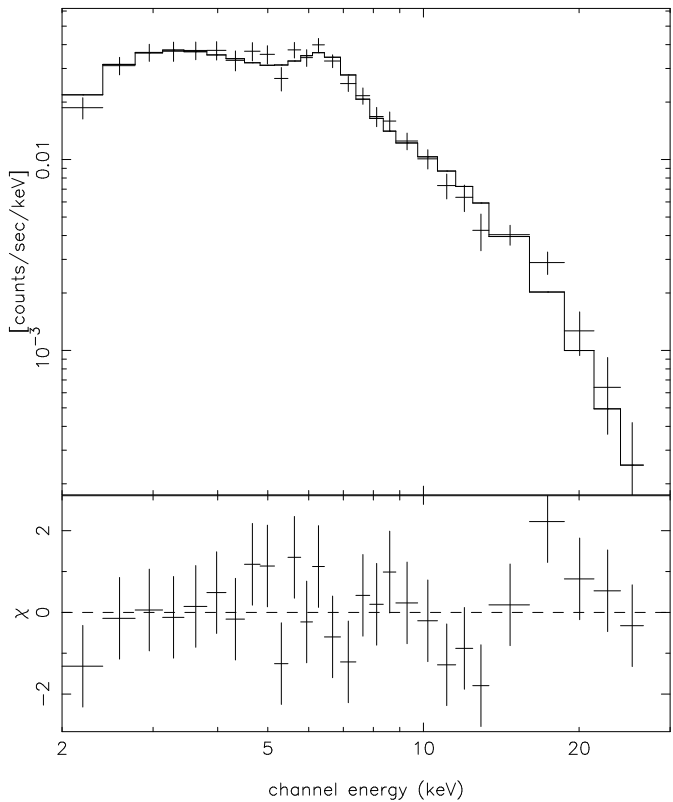}

\caption[]{Spectrum of the second flare and the fit with the Comptonized
model with black body component of temperature k$T=2.4$~keV and an
emission line at 6.4~keV. The lower panel presents the
residuals in units of sigma per channel.
\label{figwfcspe2}}
\end{figure}

We extracted a 2-26 keV spectrum of the complete Sep. 10 flare which amount 
to an elapsed time of 5.5~hr and an exposure time of 2.3~hr, and attempted 
to fit the same Comptonization model as applies to the NFI data when the 
source is $\sim20$ times
fainter. Leaving free only the normalizations of the different spectral
components, the fit was unsatisfactory with $\chi^2_{\rm r}=2.59$ for
24 dof. However, if $N_{\rm H}$ was allowed to vary, the fit became
dramatically better with $\chi^2_{\rm r}=0.89$ (23 dof) and
$N_{\rm H}=(2.5\pm0.4)\times10^{22}$~cm$^{-2}$. 
If instead the black body temperature was allowed to vary,
an even better improvement occurred with $\chi^2_{\rm r}=0.64$ (23 dof)
and k$T_{\rm bb}=2.4\pm0.2$~keV.
If the Fe-K emission 
lines at 6.68 and 6.96~keV are replaced for a single one whose energy is 
left free, the fit improves further to $\chi^2_{\rm r}=0.50$ for 23 dof. 
The line energy then is $6.39\pm0.18$~keV. This suggests a shift of the 
ionization balance to a lower degree (Fe I-XVII). We determined an average 
flux over the 3~hr duration of the peak of $5\times10^{-9}$~\ecs. The 
equivalent width of the emission line is $1.0\pm0.3$~keV. The flare spectrum 
is presented in Fig.~\ref{figwfcspe2}. 

\section{Discussion} 
\label{secdis}

We have analyzed the spectrum at basically two flux levels which correspond 
to 2-10~keV luminosities of 
$5\times10^{33}(d/{\rm 0.4~kpc})^2$~\lum\ and 
$1\times10^{35}(d/{\rm 0.4~kpc})^2$~\lum, where $d$ is the distance to \bron.
These compare to a maximum luminosity during the giant outburst
of $5\times10^{36}(d/{\rm 0.4~kpc})^2$~\lum\ 
(Smith et al. 1999) and a minimum of 
$4\times10^{32}(d/{\rm 0.4~kpc})^2$~\lum\ (see below). These 
numbers may be multiplied by a factor of 2 to obtain a rough extrapolation 
of the X-ray luminosity to the 0.4-120 keV range.
They show that, though bright, \bron\ in fact has a peak luminosity that
is at least an order of magnitude below the Eddington limit for a mass of 
the compact object that is equal to or larger than that of a canonical 
1.4~M$_\odot$ neutron star. However, one should bear in mind that the 
sampling of the ASM light curve is sparse (about ten 90~s observations per
day) and we do not know what the source did between ASM measurements.

The eye-catching feature of the {\em BeppoSAX\/} data is the strong Fe-K 
emission line complex. The line energies during the calm phase point
to a large ionization degree of the iron. Two origins for the line can be 
considered: thermal emission from a thin hot plasma, where iron atoms are 
collisionally excited, or fluorescent emission from a plasma illuminated 
by the continuum X-radiation. The thin hot plasma model gave an unsatisfactory
fit to the NFI data. Also, the fact that no difference was seen 
in equivalent width between the low- and high-flux intervals of the NFI
data seems to be at odds with a thin thermal plasma origin of the emission
line. Therefore, we believe it is more probable that we are dealing with
fluorescent emission in a photo-ionized medium. The ionization degree of
the medium is fairly high, the ionization parameter $\xi=L/nr^2$
(Kallman \& McCray 1982), where $r$ is the size of the fluorescent 
material and $n$ the atom density, is of order 10$^3$ to 10$^4$ which 
implies that the medium must be close to the source of the continuum radiation.

The 2 to 26 keV spectrum of the 2nd WFC-detected flare 
(Fig.~\ref{figwfcspe2}) shows an ionization balance of iron that is 
at much lower degrees than that of the calm emission half a year earlier.
Apparently, $\xi$ is at least two orders of magnitude 
smaller. This implies that $n$ and/or $r$ increased substantially.
Therefore, the cloud of fluorescent material must have been, during the
flare, denser and/or farther away from the source of the fluorescing 
radiation (probably near to the compact object). The same
probably applies to the giant outburst 5 days later because the iron line was
seen at an energy of 6.5~keV then (Markwardt et al. 1999b).

We do not detect K-edge absorption of iron at the appropriate energies 
of 8.8 (Fe~XXV) and 9.3 keV (Fe~XXVI). The 3$\sigma$ upper limit on the 
relative depth of an edge at 9.0~keV is 0.2 which does not appear to
be very constraining (e.g., Makishima 1986).

The equivalent width of the Fe-K line complex is relatively large.
In the WFC-detected flare in September, it was 3 to 4 times as large as
during the March NFI observation. Markwardt et al. (1999b) also
observed the Fe-K complex during the tail of the giant flare,
at a photon energy that is close to that measured with WFC six days
earlier and with an equivalent width of about 500 eV.
Model calculations of the structure of the accretion disk coronae and 
data analyses of Fe-K lines in other X-ray binaries by Vrtilek et al. 
(1993) suggest that large equivalent widths in Fe-K lines may be due to 
high (edge-on) inclination angles. This appears to be in line with the
high inclination angle inferred from the radio jets.

Hjellming et al. (1999b) reported a double-sided jet
structure from \bron\ in a VLA image taken on Sep. 16.02 UT, with 
sizes of roughly 0.25\arcsec\ in both directions. 
A subsequent VLA image taken on Sep. 17.93 only shows one of the two
jets at the same position. If $v$ is the
intrinsic velocity of the jets,
$i$ the angle between the jets and the line of sight, and $c$ the 
velocity of light, then the
angular proper motion $\mu_{\pm}$ of the approaching (--) and
receding jet (+) are given by:
\begin{eqnarray*}
\mu_{\pm} & = & (v/d) {\rm sin}i / [1\pm(v/c){\rm cos}i]
\end{eqnarray*}
(e.g., Hjellming \& Rupen 1995). We can only make a rough estimate of
$\mu_{+}$ and $\mu_{-}$. The uncertainty results from the facts that
no moving radio blobs were seen, like in GRO~J1655--40 (c.f., Hjellming \&
Rupen 1995), and that no difference was measured between $\mu_{+}$ 
and $\mu_{-}$ (this is partly due to insufficient accuracy of the
position of the optical counterpart). If we assume that the difference 
between $\mu_{+}$ and $\mu_{-}$ is less than 10\%, we find that
$(v/c){\rm cos}i<0.048$. The value for $\mu_{+}\approx\mu_{-}$
is determined by the time of jet ejection. Depending 
on whether one takes this to be at the time of the measured optical peak 
(Stubbings 1999), X-ray peak or X-ray pre-peak (see Smith et al. 1999), the 
travel time of the jets on Sep. 16.02 was between 0.3 and 1.1~d. This 
results in an allowed range for $\mu_{+}\approx\mu_{-}$ of 224 to 806 mas~d$^{-1}$.
This implies $(v/c)>0.53$. The above constraint on $(v/d){\rm cos}i$ then 
results in $i>84.8$\degr. Additionally, we can obtain a constraint on the
distance from $(v/c)<1$ and $v{\rm sin}i=\mu~d$. For $\mu=224$~mas~d$^{-1}$, 
$d<0.8$~kpc. Conversely, a distance of 0.4~kpc implies a jet ejection
on Sep. 15.4 for $v=c$ or Sep. 14.9 for $v=0.53c$

The angle $i$ is probably close to the inclination angle of the binary orbit, 
since one may expect the jets to be ejected close to perpendicular
to the binary's orbital plane. The high value of $i$ then suggests that 
at times the compact object should be eclipsed by the companion star.
Our NFI data involve the longest continuous observation ever performed 
of \bron\ while it was above the detection threshold (all {\em RossiXTE\/}
TOOs lasted shorter than 3~hr). During 19~hr no
eclipse was seen. Since the earth obscured the view to \bron\ each
{\em BeppoSAX\/} orbit, there is some uncertainty in the non-detection
of an eclipse. With this reservation, the non-detection suggests that either
the binary has an orbital period in excess of 19~hr, that the jets are
misaligned from the orbital rotation axis by more than $\sim$10\degr\ (like
seems to be the case for the micro quasar GRO~J1655--40, see Van der Hooft 
et al. 1997), or that the companion star is not an A-type main sequence star.

The X-ray flux of \bron\ behaved quite erratic. Whenever \bron\ was
detected, it exhibited at least 50\% variability on hourly time scales.
The variability is not (quasi) periodic. 
The scan observations with the PCA show that the source turned on probably
on Feb. 18 and disappeared immediately after the giant outburst on Sep. 15.
During the seven months in between, the snapshot measurements revealed fluxes 
varying at least between an upper limit of 0.4 and detections of up to 
12$\times10^{-10}$~\ecs\ (2-60 keV; 
Markwardt et al. 1999b, C.B.~Markwardt priv. comm.). The two 
WFC-detected flares show that the flux sometimes reached flux levels
one order of magnitude higher. Where does this variability come from and
why did it turn off immediately after the giant outburst?

The mass transfer from the companion star should be mediated either through
an (irregular) wind from the companion star or an accretion disk. Which one 
applies to \bron\ is unclear, given the observations thus far. The high 
variability is perhaps most easily explained by an irregular wind from 
the companion star.  Such wind accretion is very 
common among the high-mass X-ray binaries (see review by White et al. 
1995). We note that \bron\ shows similarities to CI Cam.
CI~Cam was fast (e-folding decay time 0.6~d, Harmon et al. 1998),
exhibited a bright Fe-K line (EW up to 597~eV, Orr et al. 1998),
was a radio jet source (Hjellming \& Mioduszewski 1998), and was detectable
in the 20-70 keV band (Harmon et al. 1998). The optical counterpart
of CI~Cam is a symbiotic B star with an irregular wind.
The alternative, in the accretion disk interpretation, is that
disk or thermal instabilities are continuously important. The apparent
0.3 to 0.9 day head start in the giant outburst of the optical to the
X-ray emission suggests that a disk instability may be at work during
this particular flare
which moves from the outside to the inside of the disk, like has
been seen in dwarf novae (e.g., Meyer \& Meyer-Hofmeister 1994)
and in the LMXB black-hole transient GRO~J1655--40 (Orosz et al. 1997). 
During the 5 days before the giant outburst \optic\ is
continuously 2 mag brighter than immediately after the giant outburst 
when there is hardly any X-ray emission (Kato et al. 1999). This can
be explained as optical emission from a disk which disappeared with the 
giant outburst. The alternative in the wind accretion interpretation
is that the 2 mag come from the companion star. This too is not 
unreasonable, increased levels of mass loss are likely to go hand in hand 
with such brightening.
Finally, there is the observation that the largest flare forced the
system to go in quiescence. Again, this can be explained either way.
In the accretion disk interpretation, the giant
outburst drained the disk in such a manner 
that no accretion occurs anymore. The implied
time scale of this drainage is about 10~hr. This can only be explained
if the drained mass is small though enough to induce a strong flare.

The short distance enables one to set a rather strict limit on the flux
for quiescent emission. The PCA monitoring program reveals that the 
flux went below an upper limit of 0.4$\times10^{-10}$~\ecs\ (2-60~keV) or
1~mCrab (C.B.~Markwardt, priv. comm.). Assuming the 
same spectrum holds as during our NFI observation (when the flux was 
approximately 25 times higher), the 0.4-120~keV luminosity upper limit is 
$<8\times10^{32}(d/{\rm 0.4~kpc})^2$~\lum. Unfortunately, this flux limit is
too high to say something sensible about the nature of the compact object
(see, e.g., Rutledge et al. 1999).

We reiterate the point made by Wijnands \& Van der Klis (2000) that this
transient may be an example of a separate class of fast and faint X-ray 
transients whose existence has been proposed by Heise et al. (1999) and
is characterized by peak luminosities that are two orders of magnitude
less than the typical bright soft X-ray transient and by outburst durations
of order one week instead of months. They estimate a rate of 18 such
transients per year within 20\degr\ from the Galactic center. Such
transients are easily over-seen by surveying instruments with
small duty cycles if they are at the usual few-kpc distances. It is only
because of its brightness due to its nearness that \bron\ was easily
noticed. At the distance to the Galactic center, the source would only
have had a peak flux of 30~mCrab.

It is not clear whether the compact object is a neutron star or
a black hole candidate. No X-ray characteristics that unambiguously 
identify the nature, such as type I X-ray bursts or pulsations, 
are present. Optical Doppler measurements during quiescence will 
give the best opportunity to obtain constraints on the mass of
the compact object.

\begin{acknowledgements}
We thank Frits Paerels for useful discussions. 
This research has made use of linearized and cleaned event files from
the ASI-BeppoSAX SDC on-line database and archive system. 
{\em BeppoSAX\/} is a joint Italian and Dutch program.

\end{acknowledgements}


\begin{thebibliography}{}
   \bibitem[\protect\astroncite{Boella et~al.}{1997a}]{boe97a}
      Boella G., Butler R.C., Perola G.C., et al., 1997a, 
      A\&AS 122, 299

   \bibitem[\protect\astroncite{Boella et~al.}{1997b}]{boe97b}
      Boella G., Chiappetti L., Conti G., et al. 1997b, A\&AS 122, 327

   \bibitem[\protect\astroncite{Frontera et al.}{1997}]{fro97}
	Frontera F., Costa E., Dal Fiume D., et al. 1997, A\&AS 122, 357

   \bibitem[\protect\astroncite{Garcia}{1999}]{gar99}
	Garcia M.R., McClintock J.E., Callanan P.J., 1999, IAUC 7271

   \bibitem[\protect\astroncite{Harmon}{1998}]{har98}
	Harmon B.A., Fishman G.J., Paciesas W.S., 1998, IAUC 6874

   \bibitem[\protect\astroncite{Heise}{1999}]{hei99}
      Heise J., in 't Zand J.J.M., Smith M.J., et al. 1999, in proc 
      `The Extreme Universe' (3rd Integral Workshop), eds. A. Bazzano,
      G.G.C. Palumbo, C. Winkler, Astrophys. Lett. Comm.
      38, 297

   \bibitem[\protect\astroncite{Herbig}{1975}]{her75}
	Herbig G.H. 1975, ApJ 196, 129

   \bibitem[\protect\astroncite{Hjellming}{1995}]{hje95}
	Hjellming R.M., Rupen M.P., 1995, Nat 375, 464

   \bibitem[\protect\astroncite{Hjellming et al.}{1998}]{hje98}
	Hjellming R.M., Mioduszewski, A.J., 1998, IAUC 6872

   \bibitem[\protect\astroncite{Hjellming}{1999a}]{hje99a}
	Hjellming R.M., Rupen M.P., Mioduszewski A.J., 1999a, IAUC 7254

   \bibitem[\protect\astroncite{Hjellming}{1999b}]{hje99b}
	Hjellming R.M., Rupen M.P., Mioduszewski A.J., 1999b, IAUC 7265

   \bibitem[\protect\astroncite{zand99a}{1999a}]{zan99a}
      In 't Zand J.J.M., Heise J., Bazzano A., et al. 1999a, IAUC 7119

   \bibitem[\protect\astroncite{zand99b}{1999b}]{zan99b}
      In 't Zand J.J.M., Verbunt F., Strohmayer T.E., et al. 1999b,
      A\&A 345, 100

   \bibitem[\protect\astroncite{Jager et~al.}{1997}]{jag97}
       Jager R., Mels W.A., Brinkman A.C., et al., 1997, A\&AS 125, 557

   \bibitem[\protect\astroncite{Kallman \& McCray}{1982}]{kal82}
       Kallman T., McCray R. 1982, ApJS 50, 263

   \bibitem[\protect\astroncite{Kato et~al.}{1999}]{kat99}
       Kato T., Uemera M., Stubbings R., Watanabe T., Monard B. 1999, IBVS 4777

   \bibitem[\protect\astroncite{Makishima}{1986}]{mak86}
	Makishima K. 1986, in ``The Physics of Accretion onto
        Compact Objects'', eds. K.O. Mason, M.G. Watson, N.E. White
        (Berlin: Springer Verlag), p. 249

   \bibitem[\protect\astroncite{Manzo}{1997}]{man97}
	Manzo G., Giarusso S., Santangelo A., et al. 1997, A\&AS 122, 341

   \bibitem[\protect\astroncite{Markwardt}{1999a}]{mar99a}
	Markwardt C.B., Swank J.H., Marshall F.E.., 1999a, IAUC 7120

   \bibitem[\protect\astroncite{Markwardt}{1999b}]{mar99b}
	Markwardt C.B., Swank J.H., Morgan E.H., 1999b, IAUC 7257

   \bibitem[\protect\astroncite{McCollough}{1999}]{mcc99}
	McCollough M.L., Finger M.H., Woods P.M., 1999, IAUC 7257

   \bibitem[\protect\astroncite{Mewe}{1995}]{mew95}
           Mewe R., Kaastra J.S., Liedahl D.A. 1995, Legacy 6, 16

   \bibitem[\protect\astroncite{Meyer \& Meyer-Hofmeister}{1994}]{mey94}
           Meyer F., Meyer-Hofmeister E. 1994, A\&A 288, 175

   \bibitem[\protect\astroncite{Orosz et al.}{1997}]{oro97}
           Orosz J.A., Remillard R.A., Bailyn C.D., McClintock J.E.,
           1997, ApJ 478, L83

   \bibitem[\protect\astroncite{Orr et al.}{1998}]{orr98}
           Orr A., Parmar A.N., Orlandini M., et al. 1998, A\&A 340, L19

   \bibitem[\protect\astroncite{Parmar}{1997}]{par97}
       Parmar A.N., Martins D.D.E., Bavdaz M., et al. 1997, A\&AS 122, 309

   \bibitem[\protect\astroncite{Predehl \& Schmitt}{1995}]{pre95}
   	Predehl P., Schmitt J.H.M.M. 1995, A\&A 293, 889

   \bibitem[\protect\astroncite{Rutledge et al.}{1999}]{rut99}
   	Rutledge R.E., Bildsten L., Brown E.F., Pavlov G.G.,
        Zavlin V.E. 1999, ApJ 514, 945

   \bibitem[\protect\astroncite{Samus et al.}{1999}]{sam99}
	Samus N.N., Hazen M., Williams D., Welther B., Williams G.V., 
        1999, IAUC 7277

   \bibitem[\protect\astroncite{Smith et al.}{1999}]{smi99}
	Smith D.A., Levine A.M., Morgan E.H., 1999, IAUC 7253

   \bibitem[\protect\astroncite{Stubbings}{1999}]{stu99}
	Stubbings R. 1999, IAUC 7253

   \bibitem[\protect\astroncite{Terada et al.}{1999}]{ter99}
   	Terada Y., Kaneda H., Makishima K., et al., 1999, PASJ 51,
        39

   \bibitem[\protect\astroncite{Titarchuk}{1994}]{tit94}
   	Titarchuk L. 1994, ApJ 434, 313

   \bibitem[\protect\astroncite{Van der Hooft}{1997}]{hoo97}
   	Van der Hooft F., Groot P.J., Shahbaz T., et al. 1997, MNRAS 286, L43

   \bibitem[\protect\astroncite{Titarchuk}{1994}]{par95}
   	Van Paradijs J., White N.E. 1995, ApJ 447, L34

   \bibitem[\protect\astroncite{Vrtilek}{1993}]{vir93}
   	Vrtilek S.D., Soker N., Raymond J.C., 1993, ApJ 404, 696

   \bibitem[\protect\astroncite{Wagner}{1999}]{wag99}
	Wagner R.M. 1999, IAUC 7276

   \bibitem[White, Nagase, Parmar 1995]{whi95}
      White N.E., Nagase F., Parmar A.N., 1995, in "X-ray Binaries", 
      eds. W.H.G. Lewin , J. van Paradijs, 
      E.P.J. van~den Heuvel (Cambridge University Press, Cambridge),
      p.~1
   
   \bibitem[\protect\astroncite{Wijnands \& Van der Klis}{2000}]{wijn00}
    Wijnands R., van der Klis M. 2000, ApJ 528, L93

   \bibitem[\protect\astroncite{Williams}{1999}]{wil99}
	Williams G.V. 1999, IAUC 7253

   \end{thebibliography}
\end{document}